\documentclass{IEEEtran}
\usepackage{amsmath,amsfonts}
\usepackage{algorithmic}
\usepackage{algorithm}
\usepackage{array}
\IEEEoverridecommandlockouts
\usepackage[caption=false,font=normalsize,labelfont=sf,textfont=sf]{subfig}
\usepackage{textcomp}
\usepackage{stfloats}
\usepackage{url}
\usepackage{verbatim}
\usepackage{caption}
\usepackage{graphicx}
\graphicspath{{figures/}}
\usepackage{cite}
\makeatletter
\providecommand{\@LN@col}[1]{}
\providecommand{\@LN}[2]{}
\makeatother
\usepackage{amsmath,amssymb,amsfonts}
\usepackage{orcidlink}
\usepackage{comment}
\usepackage{algorithmic}
\usepackage{textcomp}
\usepackage{tabularx}
\usepackage{titlesec}
\usepackage{graphicx}
\usepackage{threeparttable}
\usepackage{textcomp}
\usepackage{xcolor}
\usepackage{cite}
\usepackage{enumitem}
\usepackage{comment}
\usepackage{makecell}
\usepackage{moresize}
\definecolor{darkgreen}{RGB}{0,100,0}
\usepackage{multirow} 
\usepackage{hyperref}
\hypersetup{
    colorlinks=true,
    linkcolor=blue,
    citecolor=blue,
    urlcolor=blue
}
\newcommand{\rr}[1]{{\color{black} #1}}

\hypersetup{
    hidelinks,
    colorlinks=true,
    citecolor=blue,
    linkcolor=blue,
    urlcolor=blue
}
\def\BibTeX{{\rm B\kern-.05em{\sc i\kern-.025em b}\kern-.08emT\kern-.1667em\lower.7ex\hbox{E}\kern-.125emX}}

\usepackage{newtxtext,newtxmath}

\begin{document}
\title{Adaptive Inductor and Frequency Management for Integrated Power Converters}

\author{Rami~Rasheedi~\orcidlink{0009-0009-8410-8011},~\IEEEmembership{Graduate Student Member,~IEEE},
Salma~Abdelzaher~\orcidlink{0009-0003-7242-5665},~\IEEEmembership{Graduate Student Member,~IEEE},
and~Inna~Partin-Vaisband~\orcidlink{0000-0002-6399-6672},~\IEEEmembership{Senior Member,~IEEE}%
\vspace{-10pt}
\thanks{The authors are with the Department of Electrical and Computer Engineering, University of Illinois Chicago, Chicago, IL 60607, USA.}}





\maketitle

\begin{abstract}
This work presents a digitally controlled adaptive power conversion framework that dynamically co-optimizes inductor configuration and switching frequency to achieve high efficiency across a wide range of load conditions. Conventional buck converters are typically optimized for a single operating point, leading to efficiency degradation under dynamic loads due to mismatches between inductance, switching frequency, and load current. 
To address this limitation, a reconfigurable inductor architecture is proposed. With this approach, the effective inductance and switching frequency are adjusted at runtime with on-chip digital controller.
An end-to-end design methodology is developed, integrating physics-based analytical modeling, finite element method (FEM) simulations, and an optimization framework to determine preferred operating points across the load range. The resulting configurations are stored in a lookup table and implemented in real time through a low-overhead control scheme. The system explicitly accounts for both inductor and switching device losses, enabling co-optimization of conduction and switching losses within current ripple and continuous conduction mode (CCM) constraints.
To ensure robust operation under fast load transients, a hysteresis-based control
strategy is introduced to mitigate temporary CCM violations caused by sensing
latency. An entry-margin condition is derived under which continuous conduction is guaranteed for a bounded load slew rate. Simulation results demonstrate a 4.5 percentage-point improvement in average efficiency and up to a 32\% reduction in average power loss relative to fixed
configurations, highlighting the effectiveness of adaptive inductance-frequency management in integrated power delivery systems with dynamic load profiles.
\end{abstract}


\vspace{-5pt}

\section{Introduction}
\IEEEPARstart{P}{ackage}-integrated power delivery systems must deliver high current density while maintaining high conversion efficiency~\cite{radhakrishnan2021power,1, rasheedi2026package}. Within these converters, the inductor is central to energy storage and filtering but is also a primary loss contributor, through both DC winding resistance and AC (core, skin, and proximity) effects; at high load the $I^2R$ conduction loss dominates owing to its quadratic current dependence, while AC losses depend more weakly on load~\cite{8385217, 9124544}. Inductor design is therefore a trade-off: lowering DC resistance favors fewer turns or thicker conductors, which reduces inductance, whereas raising inductance suppresses ripple and AC loss but increases winding resistance and size~\cite{9816521, 9097695}.

Switching devices are likewise a major loss source~\cite{ECTC-salma}: conduction loss grows quadratically with current, while switching loss grows with both current and switching frequency. Lowering the frequency reduces switching loss but raises ripple and inductor loss, so a minimum-loss design must jointly account for switch and inductor losses. In a conventional buck converter, inductance and frequency are fixed to maximize efficiency at a nominal operating point~\cite{alvarez2019open}. Real workloads in edge and high-performance computing, however, are strongly time-varying~\cite{radio, rasheedi202528, 10509290, kong2025vertical}, so a fixed design is suboptimal away from that point---wasting conduction loss at heavy load and switching loss at light load---motivating runtime adaptation of the converter.

A broad class of \emph{dynamically adjustable} converters combats this by adapting
operation to the instantaneous load; the discussion is framed around the buck
topology for its ubiquity in point-of-load regulation, though the principles
generalize to most inductor-based converters. Reported techniques vary the
switching frequency with load~\cite{villar2005efficiency},
hop between PWM, PFM, and burst/pulse-skipping modes at light
load~\cite{sahu2007adaptivepfm}, shed or add phases in multiphase
designs~\cite{su2013phaseshedding}, or adapt dead time and gate
drive at the device level~\cite{yousefzadeh2006sensorless,
musunuri2005width}. Crucially, all of these adapt only the
\emph{switching and control domain} while the inductor remains
fixed~\cite{trescases2011survey, maksimovic2004impact}. An inductor sized for one
operating point is therefore mismatched at others, so these methods typically
benefit only a single load extreme and their gains vanish once inductor conduction
loss dominates, while adding sensing, mode-detection, and control-loop
complexity.

These observations motivate adapting the magnetic element itself. A reconfigurable
inductor introduces the inductance as an additional, load-dependent degree of
freedom, directly addressing the inductance-versus-load mismatch and the inductor
losses that control-domain techniques leave untouched; because it acts on an
orthogonal part of the converter, it is complementary to any of the above.
Variable inductors have been explored but are dominated by magnetically-biased
saturable-core devices, in which a DC bias drives a gapped ferrite toward
saturation to vary its permeability~\cite{azim2024review, he2023review,
saeed2021dab, marques2024dab, beraki2021powertrain, nacar2025led}. These are
applied almost exclusively to isolated resonant (LLC, DAB) and PFC stages---to
track a resonant point, extend zero-voltage switching, or reduce circulating
current~\cite{saeed2021dab, jo2026llc, choi2023crm, teng2023llc}. Such devices are
intrinsically board-level, magnetic-core parts requiring auxiliary windings or an
entire auxiliary stage---overhead that can erode or even negate the efficiency
benefit~\cite{saeed2021dab, marques2024dab}---and their tuning is continuous, slow,
and hard to control precisely, while the few discrete alternatives are only
two-state~\cite{azim2024review, he2023review}. Most importantly, all adapt
inductance \emph{alone}: none jointly optimizes inductance and switching frequency
against a combined inductor-plus-switch loss model, nor enforces explicit
current-ripple and CCM constraints, in a non-isolated point-of-load buck.

To address this challenge, a digitally controlled adaptive converter incorporating a reconfigurable inductor is proposed, enabling dynamic load-dependent adjustment of (i) the effective inductance through real-time reconfiguration of the inductor turns, and (ii) the converter switching frequency. The primary objective is to minimize the total power loss over a target load-current range, including both inductor and switching losses, while satisfying inductor current ripple and form factor (i.e., embeddable) constraints.

This work specifically targets package-level embedded power delivery. As HPC processors continue to demand kiloampere-class current delivery, maximizing efficiency at high load has become increasingly important. While state-of-the-art PCB-level converters achieve very high efficiencies by employing comparatively large discrete magnetic components, such designs are incompatible with dense package-level integration, making high-load efficiency optimization of compact embedded converters a key design challenge.

By co-optimizing inductance and switching frequency, the proposed approach enables efficient converter operation over a wide load range, overcoming the limitations of conventional fixed-parameter designs~\cite{bharath2021integrated, DPMIH44, avula2024design, bellaredj2019design}. 
%
%
The main contributions of this work are:
\begin{itemize}[leftmargin=*]
    \item \textbf{A digitally controlled adaptive power conversion framework} that jointly tunes effective inductance and switching frequency in real time using a reconfigurable inductor and a compact on-chip controller, enabling efficient operation across a wide range of load conditions.

    \item \textbf{An analytical system-level co-design and optimization approach} that captures the coupled impact of inductance configuration and switching frequency on both inductor and switching losses. The framework determines near-optimal operating points across the load range under ripple and CCM constraints, and maps them to a low-overhead lookup-table-based runtime policy.

     \item \textbf{A hysteresis mechanism and a closed-form entry-margin condition} that supports stable operation under dynamic loads and guarantees continuous conduction under bounded load slew rate.

    \item \textbf{End-to-end system evaluation} demonstrating a 4.5 percentage-point improvement in average efficiency and up to a 32\% reduction in average power loss relative to fixed configurations, validating the benefits of coordinated inductance and frequency adaptation.

\end{itemize}

The remainder of this paper is organized as follows. 
The proposed adaptive converter architecture and its operational principles are presented in Section~\ref{Configurable-Converter}. The analytical optimization framework and the associated design methodology are described in Section~\ref{Workflow}. Analytical and circuit-level validation of the proposed approach, including comparisons with conventional fixed-parameter and frequency-adaptive buck converters, are provided in Section~\ref{Results_Discussion}. Finally, the paper is concluded and the key findings are summarized in Section~\ref{Conclusion}.

\vspace{-5pt}

\section{Proposed Adaptive Converter}
\label{Configurable-Converter}
\begin{figure}[tb]
    \centerline{\includegraphics[width=\columnwidth]{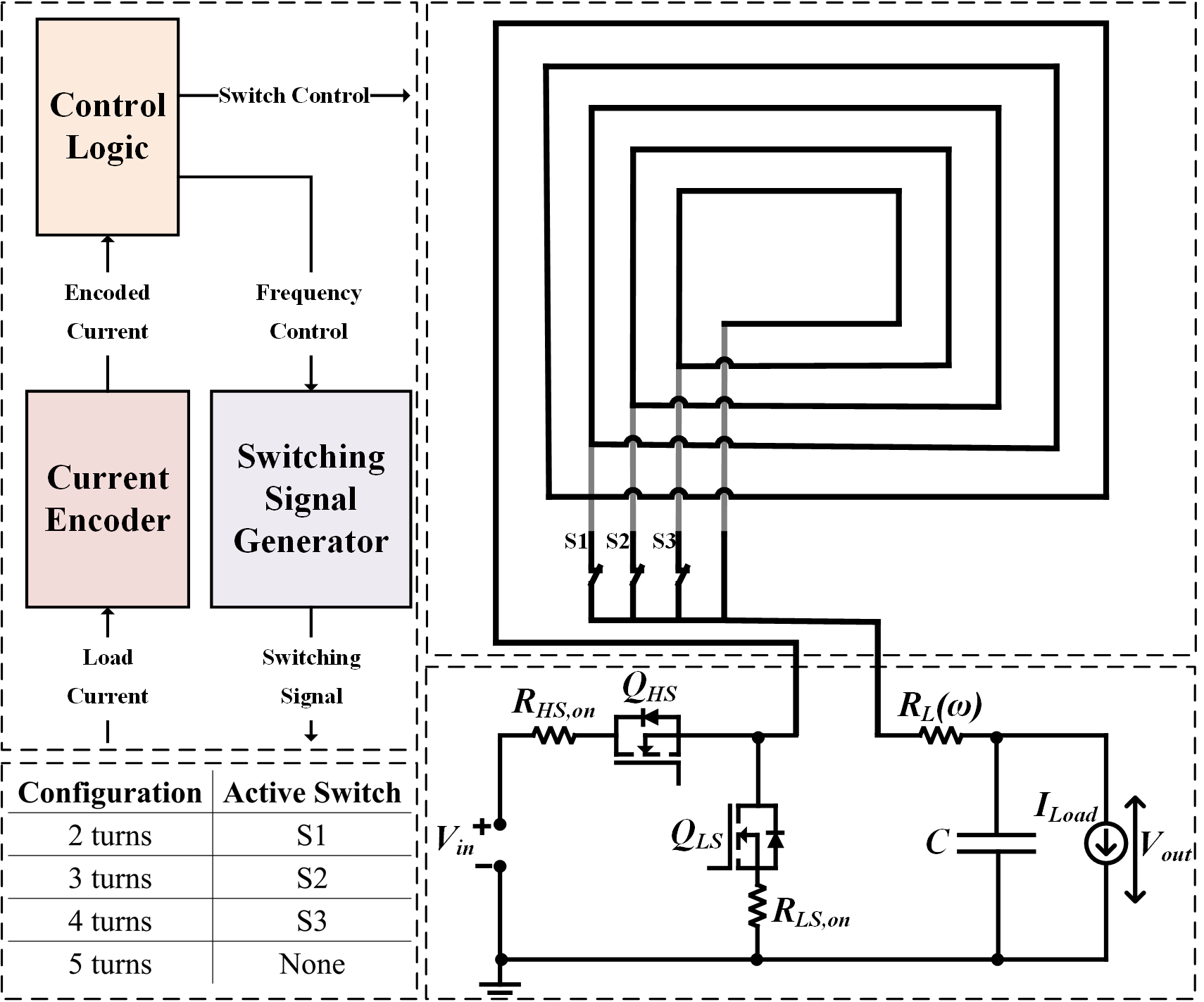}}
    \caption{The proposed adaptive system architecture, comprising a reconfigurable inductor and switching-frequency control, with key components identified.}
    \vspace{-15pt}
    \label{Adaptive_system}
\end{figure}

The proposed adaptive system, as illustrated in Fig.~\ref{Adaptive_system}, consists of a reconfigurable inductor, a switching-frequency control mechanism, and a buck converter. The approach can be applied to any inductor-based converter topology. The buck converter is adopted solely for clarity and simplicity of demonstration. The reconfigurable inductor is implemented as a planar, air-core square spiral, based on the embedded spiral inductor topology and analytical models in~\cite{rasheedi2024embedded, ramiectc}. 

The square-spiral geometry is adopted for its high inductance density and area
utilization in a package-embedded process, and because its concentric turns can
be conveniently brought out as intermediate taps.
The winding is divided into sub-inductors connected in series, and the nodes between
them are exposed as taps; a small network of configuration switches then selects
which tap is connected into the converter, so that the effective number of active
turns---and hence the effective inductance---can be adjusted at runtime (the
specific realization, comprising four series sub-inductors and switch-selected
taps spanning two to five turns, is detailed in
Section~\ref{Results_Discussion} and Table~\ref{tab:inductor_params}). An air
core is chosen deliberately: it eliminates core and hysteresis losses, avoids
saturation, and yields a load-independent, linear inductance that is essential
for predictable operation across the wide current range. The reconfiguration
principle itself is architecture-agnostic---any multi-turn structure with
accessible intermediate taps, such as circular, hexagonal, or octagonal spirals,
stacked multi-layer spirals, solenoids, or toroids, could be reconfigured in the
same manner---so the square-spiral air-core device is one convenient realization
rather than a requirement of the method.

Active turn count as a function of load is optimized offline. The largest count (five) serves the lightest loads, where high inductance holds the ripple within tight limits and maintains CCM despite the small DC current, and the higher winding resistance is tolerable since $I^{2}R$ loss is small there. As load rises, the optimizer reduces the turns to four, three, and finally two to cut the now-dominant conduction loss, while raising the switching frequency to offset the reduced inductance and keep ripple bounded. The bounds are physical: the maximum count is limited by growing DC and AC (skin/proximity) losses, footprint, and reduced self-resonant frequency, while the minimum is set by the ripple/CCM constraints, since too few turns would violate them even at the highest frequency and heaviest load.

This approach also entails limitations. Each active configuration switch lies in the inductor current path, adding on-resistance and parasitic capacitance (Table~\ref{tab:converter_loss}). More importantly, the adaptation granularity is set by the total number of switches, since each added tap adds series resistance, parasitics, and---most critically---silicon and interconnect area. Thus, only a modest set of configurations (four here, see Fig.~\ref{Adaptive_system}) is supported. Exposing the taps of a multi-layer embedded spiral also requires extra vias and closely co-integrated switches. In addition, operation at the minimum turn count concentrates the DC current in fewer turns, imposing a thermally determined lower bound on the conductor cross-section. Together, these electrical, area, fabrication, and thermal factors set the practical range and resolution of the reconfigurable inductance.

Although demonstrated on a buck converter, the proposed approach is applicable to any inductor-based
converter. 
For each load region, the offline optimizer selects the
effective inductance 
and switching frequency to minimize total
loss subject to the topology-specific constraints, e.g., meeting ripple and CCM constraints while minimizing DCR
in filter-inductor and hybrid/series-capacitor buck converters, or maintaining resonance to preserve ZVS and minimize circulating current in LLC/DAB
converters~\cite{he2023review, azim2024review, jo2026llc, saeed2021dab}. 
The buck converter is adopted solely for clarity and simplicity of demonstration.

\begin{figure*}[t]
    \centerline{\includegraphics[width=\linewidth]{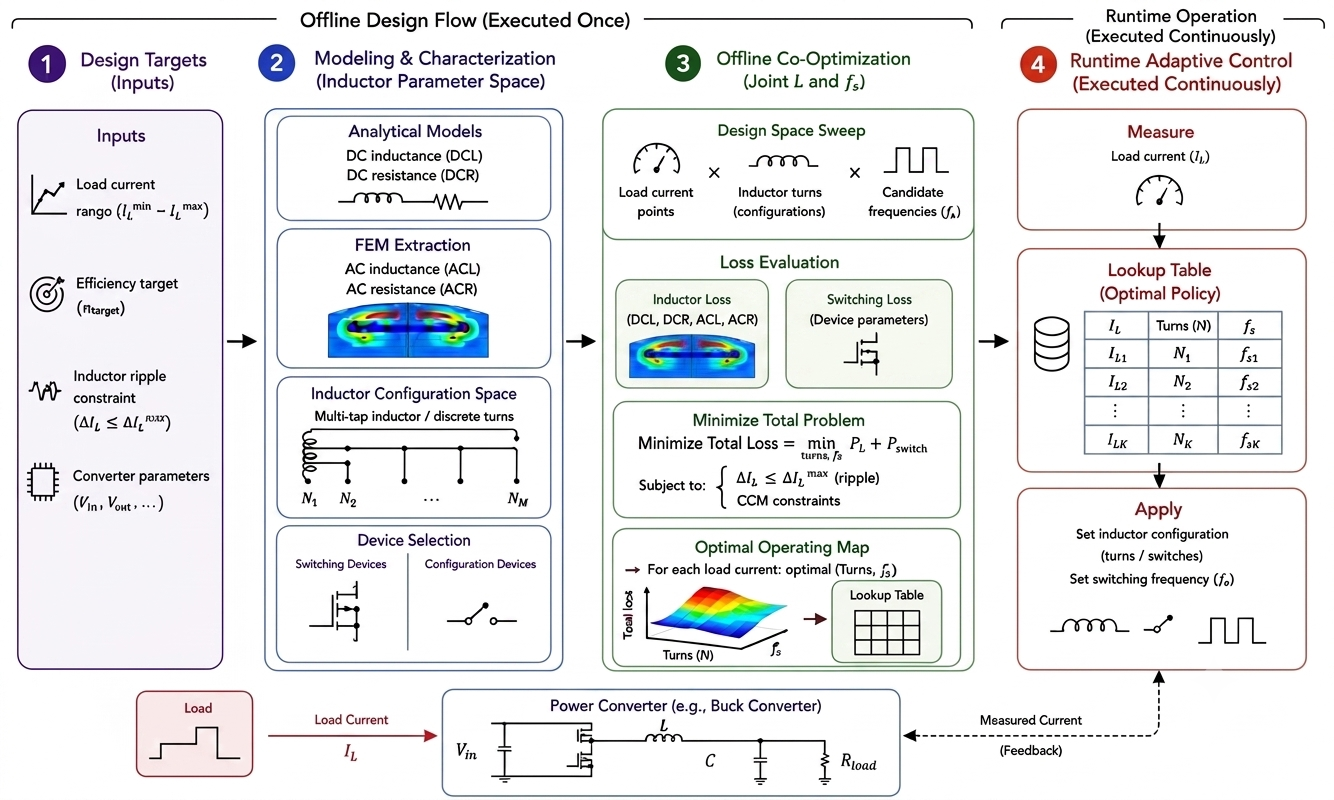}}
    \caption{Comprehensive design and optimization framework of the proposed adaptive converter, highlighting the integration of analytical modeling, FEM simulations, offline loss optimization, and real-time control via a lookup table.}
    \vspace{-15pt}
\label{workflow}
\end{figure*}

The controller digitizes the sensed load current and uses a lookup table to determine the effective number of inductor turns and the converter switching frequency. As a result,
the inductor current ripple, $\Delta I_L = D(V_{in}-V_{out})/(L f_s)$, is held within its limit while the conduction and switching losses of both the inductor and the power devices are jointly minimized. 

\begin{table}[tb]
    \centering
    \vspace{5pt}
    \caption{Current-Dependent Loss Scaling in Fixed and Adaptive Converters.  }
    \label{tab:converter_loss}
    \renewcommand{\arraystretch}{1.2}
    \resizebox{\columnwidth}{!}{%
    \begin{tabular}{l|c|cl}
        \hline
        \textbf{Loss Component} & \textbf{Fixed} & \multicolumn{2}{l}{\textbf{Proposed Adaptive}} \\
        \hline
        Current load $I_L$ & $\uparrow$ & $\uparrow$ & \\
        \hline
        Switch DC loss & $\uparrow\uparrow$ & $\uparrow\uparrow$ & same in both \\
        Switch AC loss & $\uparrow\uparrow$ & $\uparrow$ & tuned via $f_s$ \\
        Inductor DC loss & $\uparrow\uparrow\uparrow$ & $\uparrow\uparrow$ & reduced via $\# turns$ \\
        Inductor AC loss & $\uparrow\uparrow$ & $\uparrow$ & tuned via $L, f_s$ \\
        Reconfiguration network & --- & $\uparrow$ & overhead \\
        \hline
        Total loss & $\uparrow\uparrow\uparrow$ & $\uparrow$ & net reduction \\
        \hline
        \multicolumn{4}{l}{\ssmall $^*$Arrow multiplicity qualitatively indicates the strength of dependence on load current.}
        \vspace{-10pt}
    \end{tabular}}
    \vspace{-5pt}
\end{table}

\vspace{-5pt}

\section{Optimization and Workflow}
\label{Workflow}
The design workflow adopted in this work is illustrated in Fig.~\ref{workflow}. 
The proposed framework leverages existing analytical models~\cite{ramiectc, rasheedi2024embedded, ECTC-salma} as part of the inductor and converter modeling and characterization phase. An example of target specifications, including conversion step, efficiency, load current range, and allowable ripple constraints, as fed into the framework, is shown in Table~\ref{tab:system_params} (see \textit{Design Targets} in Fig.~\ref{workflow}). These specifications are used throughout the paper, without loss of generality, to demonstrate the proposed framework.

\begin{table}[t]
    \centering
    \renewcommand{\arraystretch}{1.3}
    \caption{Design Targets and Analytically-Determined Parameters and Device Choices.}
    \label{tab:system_params}
    \resizebox{\columnwidth}{!}{%
    \begin{tabular}{l|c||l|c}
        \hline
        \textbf{Parameter} & \textbf{Value} & \textbf{Parameter} & \textbf{Value} \\
        \hline\hline
        \multicolumn{4}{l}{\textit{Target Performance}} \\
        \hline
        \textbf{Voltage Conversion} & $6\to1$ V & \textbf{Load Current} & 1--6 A \\
        \hline
        \textbf{Target Efficiency} & 84\% & \textbf{Max.\ Ripple} & $50\%\,I_{\text{Load}}$ \\
        \hline\hline
        \multicolumn{4}{l}{\textit{Analytically Determined Design Parameters}} \\
        \hline
        \textbf{\# Turns} & 2--5 & \textbf{HS Switch} & EPC2040 \\
        \hline
        \textbf{Thickness} & 200 \textmu m & \textbf{LS Switch} & EPC7002 \\
        \hline
        \textbf{Outer Diameter} & 9.4 mm & \textbf{Configuration Switch} & EPC2204A \\
        \hline
        \textbf{Turn Separation} & 20 \textmu m & \textbf{Switch Frequency (MHz)} & 7.4, 8.7, 10 \\
        \hline
        \textbf{Turn Width} & 225 \textmu m & \textbf{Core Type} & Air\\
        \hline
    \end{tabular}}
\end{table}

These targets are used to (i) guide physics-based analytical models~\cite{ramiectc, rasheedi2024embedded} for estimating the initial inductor dimensions required to achieve the desired inductance, and (ii) inform the selection and sizing of switching devices, ensuring that both conduction and switching losses are accounted for from the early stages of the design (see \textit{Modeling \& Characterization} in Fig.~\ref{workflow}).
Following this initial phase, finite element method (FEM) simulations are employed to accurately capture the frequency-dependent characteristics of the inductor, including parasitics and AC loss effects under the anticipated operating conditions. The extracted inductor switching device parameters, are then used as inputs to an exhaustive search  (see \textit{Offline Co-Optimization} in Fig.~\ref{workflow}).
All feasible combinations of switching frequency and inductor configurations are systematically evaluated across the target load range. For each operating point, the configuration that minimizes the total power loss, including both inductor and switching device losses, while satisfying the inductor current ripple constraints, is identified. The loss is evaluated using accurate analytical models for both the inductor and the switching devices~\cite{ECTC-salma}, enabling a comprehensive and consistent design space exploration.
The resulting optimal configurations are stored in a lookup table implemented within the on-chip controller. At runtime, the controller references this table to dynamically select the preferred inductor configuration and switching frequency in response to changing load conditions, thereby maintaining near-optimal efficiency across the entire operating range.

\vspace{-5pt}

\section{Results and Discussion}
\label{Results_Discussion}
The proposed adaptive converter is developed and validated using analytical modeling, FEM simulations, and circuit-level simulations. Physics-based analytical models are used for converter optimization, loss estimation, and switching-frequency selection. The reconfigurable inductor is characterized using FEM simulations in ANSYS Q3D to extract inductance, resistance, and parasitic effects. The complete adaptive converter, including the power stage and control circuitry, is implemented and simulated in Cadence Virtuoso using Spectre and Verilog-A models for final circuit-level validation. 

Analytical steady-state and transient performance analysis is presented in subsections~\ref{sec:analytical_steady_state} and~\ref{sec:analytical_transient}, respectively. Circuit-level simulation results of the reconfigurable converter are reported in subsection~\ref{sec:circuit_simulation} and compared with state-of-the-art converters in subsection~\ref{sec:sota}.

\begin{figure}[tb]
\centering
\includegraphics[width=\columnwidth]{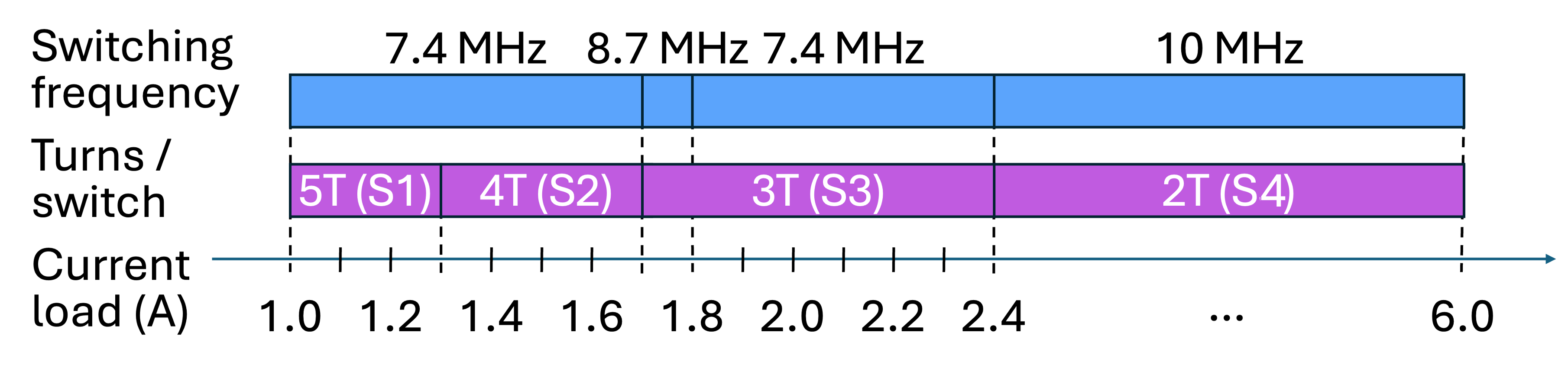}
\caption{Optimal switching frequency and inductor turn
configurations across different load current ranges, as determined by the analytical optimization for performance design targets in Table~\ref{tab:system_params}.}
\vspace{-15pt}
\label{fig:freq_and_turns}
\end{figure}

\vspace{-5pt}

\subsection{Analytical Steady-State Performance vs. Load Current}\label{sec:analytical_steady_state}

Inductor and switching-device parameters derived from the analytical optimization are summarized in Table~\ref{tab:system_params}, and the resulting switching-frequency and inductor-turn adaptation across the load range is shown in Fig.~\ref{fig:freq_and_turns}. To evaluate the proposed approach, buck converters employing conventional fixed inductors with 2--5 turns are designed and implemented in Cadence Virtuoso. Each fixed-turn converter is individually optimized to achieve a maximum efficiency of $\sim$84\% and, like the proposed system, is permitted to adapt its switching frequency with load. Consequently, the fixed-turn baselines already represent frequency-adaptive converters and differ from the proposed system only in that their inductance remains fixed. Thus, comparison with these baselines isolates the additional benefit of inductance reconfiguration rather than that of switching-frequency adaptation alone.

\begin{figure}[tb]
    \centering
    
    \subfloat[]{
        \includegraphics[width=\columnwidth]{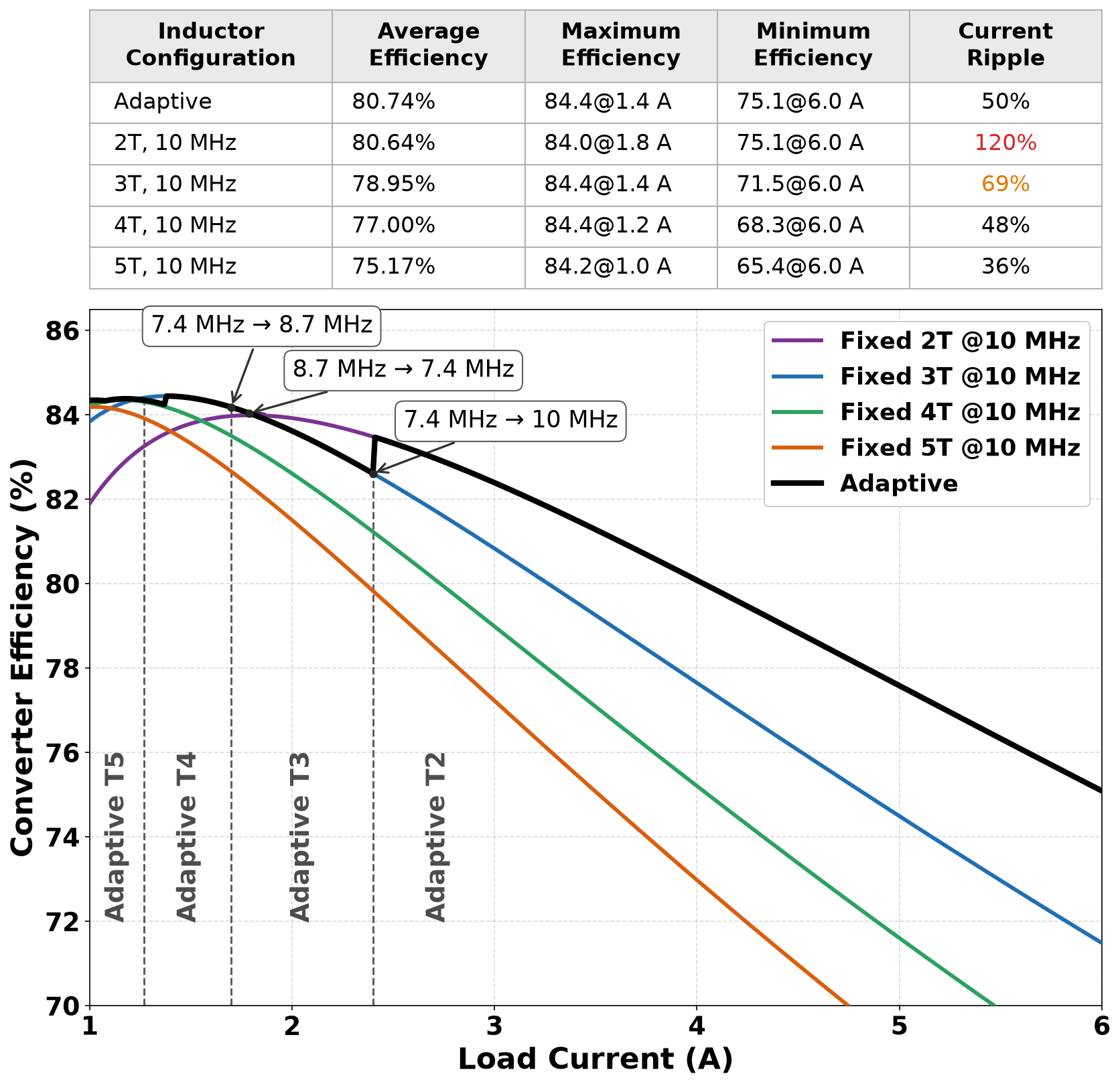}
    }
    \vspace{-10pt}
    \hfill
    \subfloat[]{
        \includegraphics[height=0.37\columnwidth]{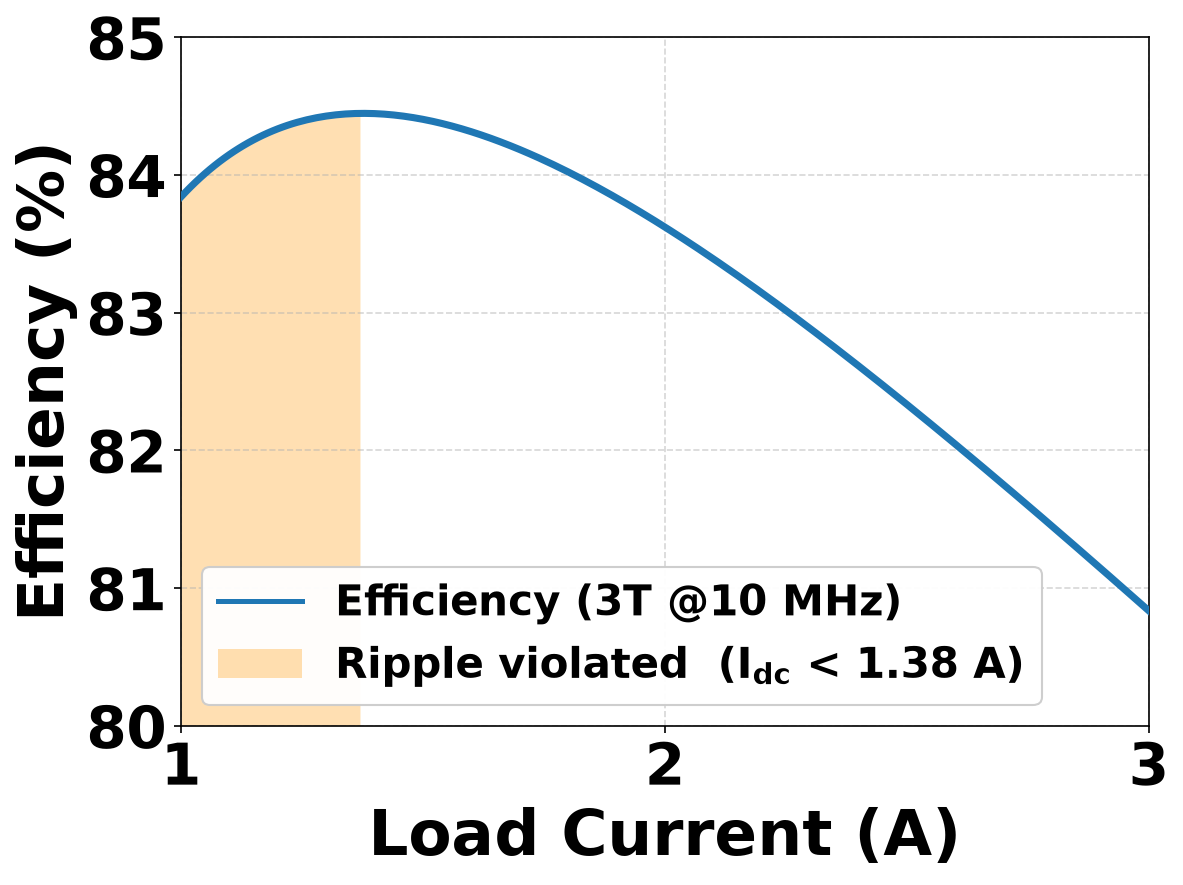}
    }
    \hspace{-10pt}
    \subfloat[]{
        \includegraphics[height=0.37\columnwidth]{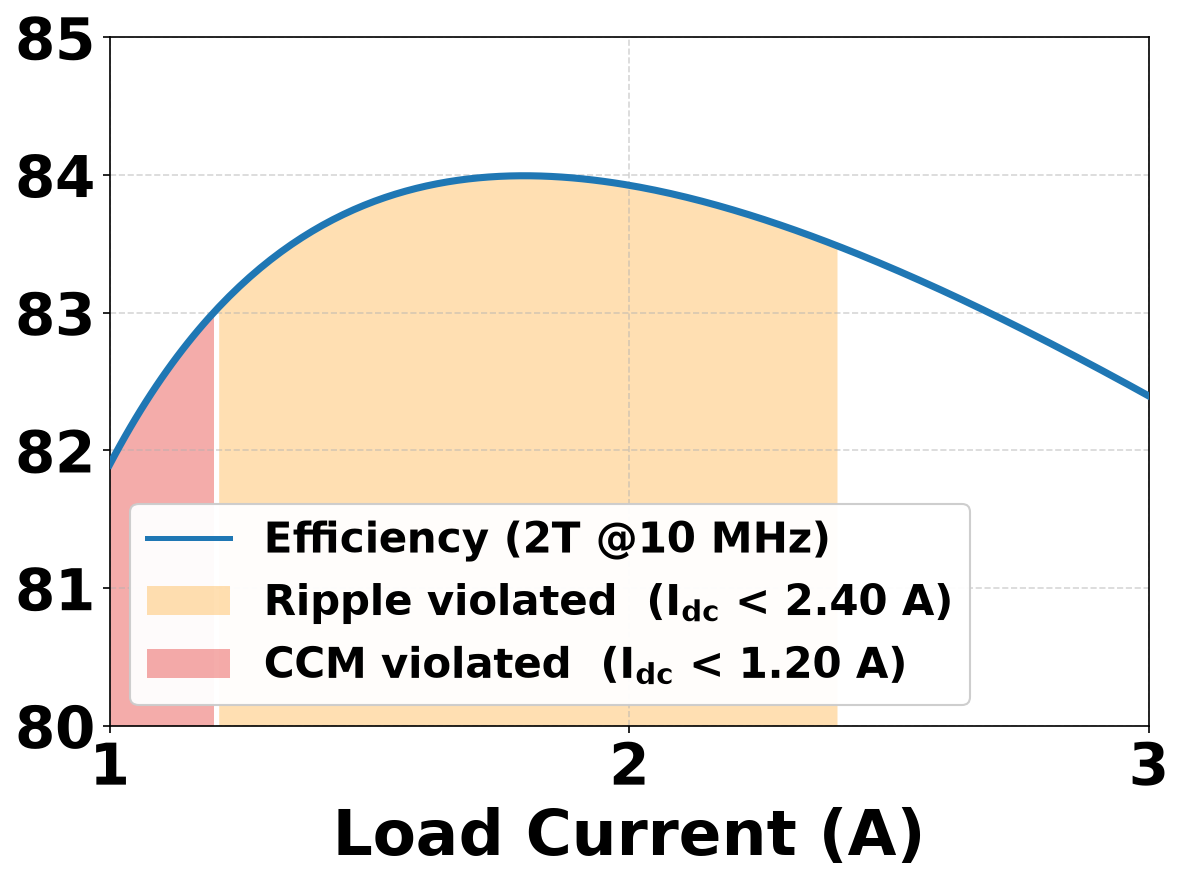}
        \hspace{-15pt}
    }
    \caption{Converter performance for adaptive and fixed inductor configurations. The adaptive scheme dynamically selects the optimal number of turns and switching frequency, achieving near-envelope performance across the load range. (a) Overall efficiency comparison, illustrating adaptively selected switching frequency and turn configurations. (b) Detailed view of the 3-turn (T3) configuration, highlighting violation of the 50\% inductor current ripple constraint at low load currents. (c) Detailed view of the 2-turn (T2) configuration, showing violations of both CCM operation and ripple constraints.}
    \vspace{-15pt}
    
    \label{generic_results_total_a}
\end{figure}

Performance comparison between the proposed adaptive converter and conventional fixed-inductor designs is presented in Fig.~\ref{generic_results_total_a}. While all configurations achieve similar peak efficiency, the adaptive system consistently outperforms the fixed 3-turn (3T), 4-turn (4T), and 5-turn (5T) designs in average efficiency, current ripple, and/or CCM operation over a broader load range. Although the fixed 2-turn (2T) configuration achieves comparable efficiency at high load, it violates the ripple constraint below 2.4~A and exits CCM below 1.2~A, rendering it unsuitable.

\begin{figure}[tb]
\centering
\includegraphics[width=\columnwidth]{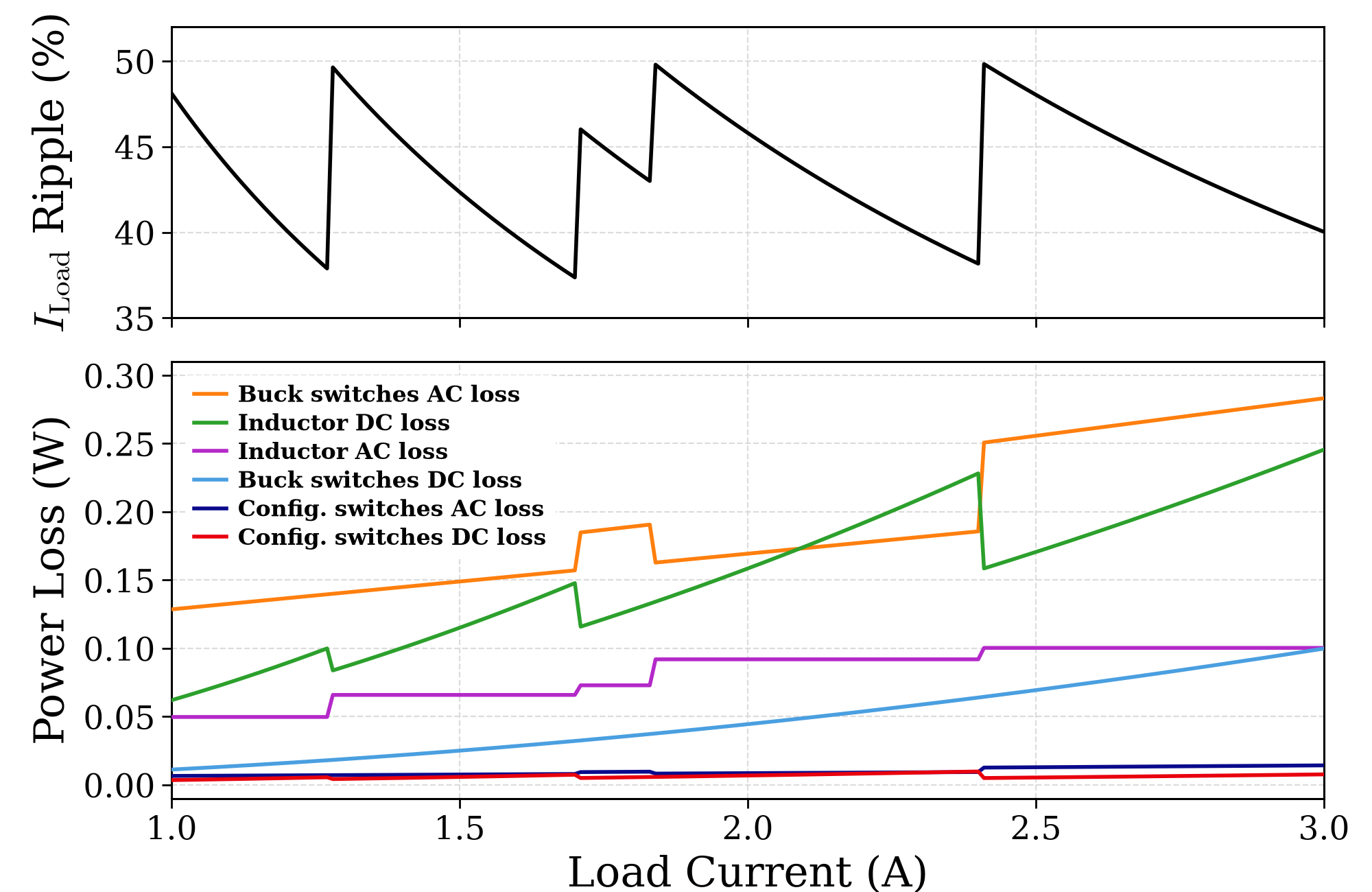}
\caption{Detailed loss breakdown of the proposed adaptive converter across the load current range. The top panel shows the resulting load-current ripple, which remains within the 50\% budget, while the bottom panel shows the contributions of inductor, buck switching, and configuration switch losses.}
\vspace{-10pt}
\label{generic_results_total_d}
\end{figure}

Light-load behavior of the fixed 2T and 3T configurations is highlighted in Figs.~\ref{generic_results_total_a}(b)--(c). The ripple constraint is violated below 2.4~A and 1.37~A, respectively, even at the maximum switching frequency. Furthermore, the 2T configuration fails to maintain CCM. Consequently, neither configuration satisfies the design requirements or is suitable for practical implementation.
The loss breakdown of the proposed adaptive system is shown in Fig.~\ref{generic_results_total_d}, illustrating how the individual loss components are balanced through dynamic adjustment of the converter configuration to minimize the total power dissipation across the load range \rr{while keeping the load-current ripple within the 50\% budget, as shown in the top panel}.

\rr{The breakdown is plotted over 1--3~A because all turn transitions occur in this range (5T$\to$4T, 4T$\to$3T, and 3T$\to$2T near 1.3~A, 1.7~A, and 2.4~A, respectively); above 3~A the converter remains in the 2T configuration. The discrete dip in a loss component at each transition---e.g., the 3T$\to$2T switch lowers the winding DCR and thus, the inductor conduction loss---shows how the losses rebalance as the configuration adapts to load.}

In the targeted package-level embedded power-delivery implementation, inductor DC losses account for up to 30\% of the total power loss because of the deliberately compact 200~\textmu m air-core winding and the $\leq$50\% ripple budget. This reflects a fundamental limitation of compact embedded inductors rather than a shortcoming of the proposed framework, which is independent of the specific converter implementation and can be readily applied to higher-efficiency converter designs.

\rr{For the best valid fixed configuration (4T at 10~MHz), adapting the switching frequency alone lowers the average power loss by 4.5\%, raising the average efficiency from 75.3\% to 76.3\%. Reconfiguring the inductance in addition reduces the average loss by a further 21.8\%, reaching 79.6\%. Overall, the average power loss is reduced by 25.4\%, and the efficiency is improved by between $+1.7$ percentage points at light load and $+7.5$ percentage points at 6~A, averaging $+4.3$ percentage points across the 1--6~A range. Inductance reconfiguration thus contributes the dominant share of the improvement, its benefit growing at high load as the inductor conduction loss becomes dominant.}

Inductance reconfiguration is complementary to dynamic techniques such as phase shedding, PWM/PFM/burst mode, adaptive dead time, and adaptive gate drive. These techniques act primarily in the switching and control domains, whereas inductance reconfiguration addresses the inductance-versus-load mismatch together with the associated inductor $I^{2}R$ and AC losses. Consequently, their benefits are additive. The magnitude of the improvement depends on the inductor contribution to the total converter loss and therefore cannot be generalized across all converter implementations. A quantitative comparison with each dynamic technique is beyond the scope of this work because several assume different converter topologies or target only a specific operating regime.

\vspace{-5pt}

\subsection{Analytical Transient Load Validation and Hysteresis-Based DCM Mitigation}\label{sec:analytical_transient}
\begin{figure}[tb]
\centering
\includegraphics[width=\columnwidth]{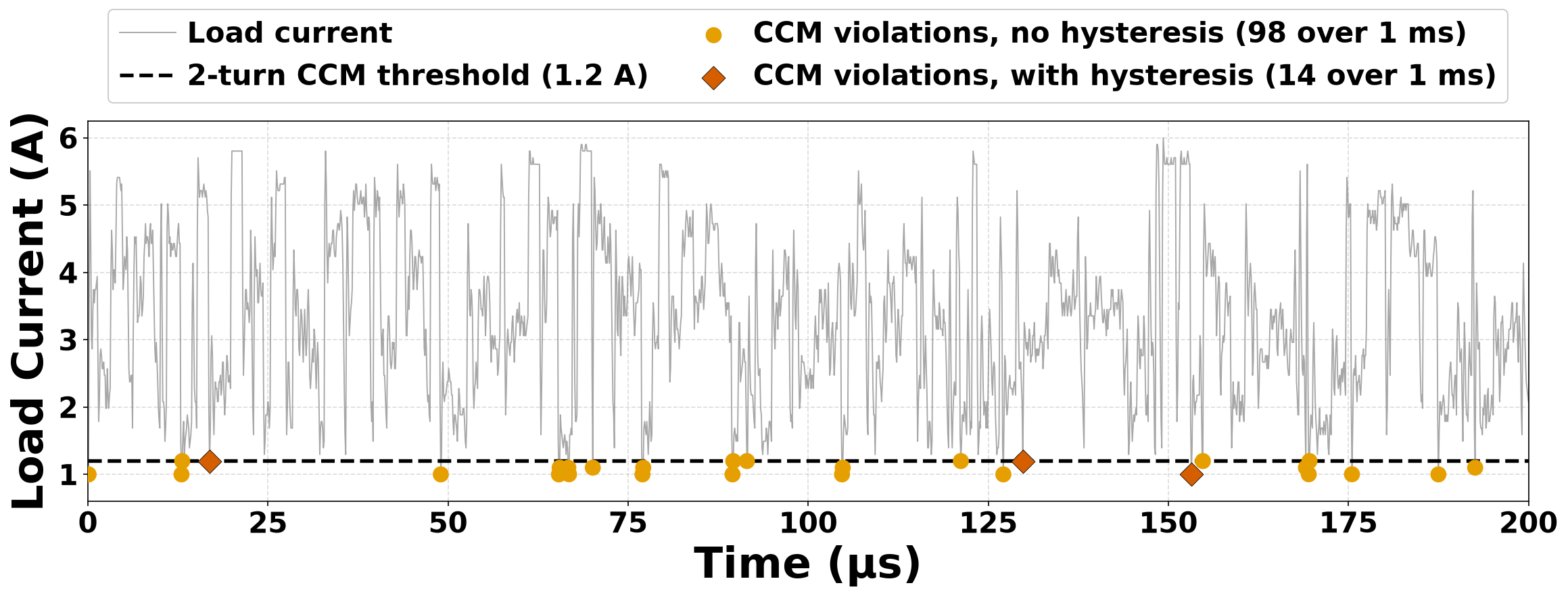}
\caption{Time-domain load current profile illustrating CCM violations for the adaptive system under normal and hysteresis-based operation, along with the 2-turn CCM current threshold.}
\vspace{-15pt}
\label{ccm_total_b}
\end{figure}

To evaluate robustness under highly dynamic operating conditions, the controller is validated using a fast-varying load-current profile representative of bursty workload behavior in modern AI accelerators, GPUs, and heterogeneous edge-computing systems.
The load current is synthesized by aggregating current loads of multiple low-dropout regulators (LDOs)~\cite{ldo} operating concurrently over time. The resulting current profile spans the target current range of interest while capturing dynamic variations encountered in practical applications. These data are used to evaluate the transient performance of the converter under time-varying load conditions. A representative segment of the synthetic load-current profile is shown by the gray trace in Fig.~\ref{ccm_total_b}. 

At a practical 1~MHz ADC sampling rate~\cite{ad4691_datasheet}, fast sub-\textmu s load transients may not be detected immediately, causing the converter to operate briefly ($<$1~\textmu s) in a suboptimal configuration. This might result in a small efficiency penalty and, more importantly, a temporary CCM violation. This occurs when a low-turn configuration (e.g., 2-turn) is latched at high load and the load current subsequently drops sharply before the controller can increase the number of turns, momentarily forcing discontinuous conduction mode (DCM).

To mitigate this, hysteresis is introduced into the turn-selection logic: the 3T$\rightarrow$2T transition uses an elevated entry threshold of 5~A while the load current is increasing, and reverts to the nominal 2.41~A threshold as the load current decreases. This asymmetric band avoids latching the low-inductance 2-turn state except at the top of the load range, preventing rapid toggling and suppressing transient CCM violations.

\begin{figure}[tb]
\centering
\includegraphics[width=\columnwidth]{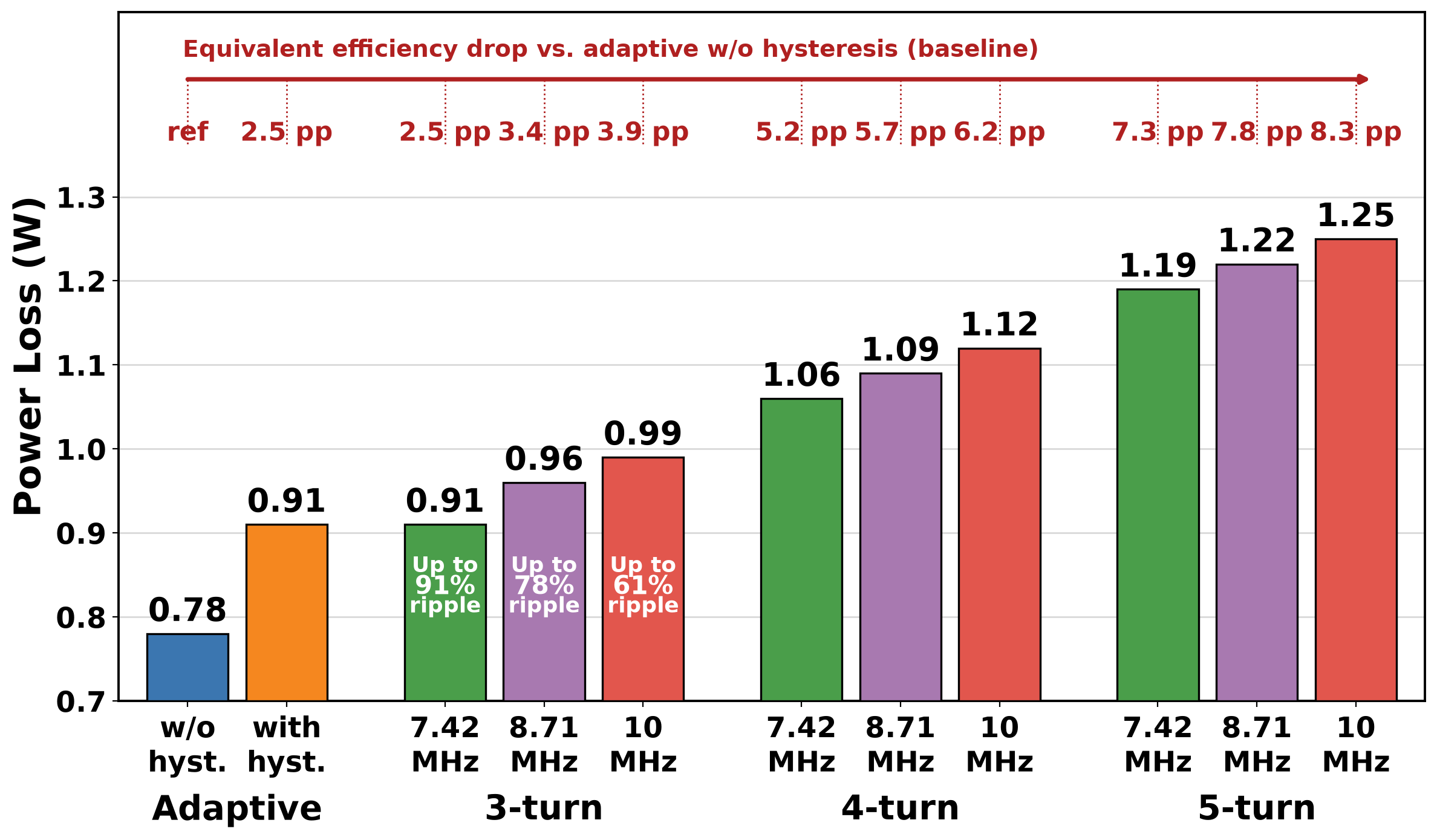}
\caption{Average power loss over a 1 ms interval for the proposed adaptive system and conventional fixed-inductor configurations (3-5 turns), highlighting the impact of adaptive inductance and switching frequency selection. Configurations that violate the 50\% ripple threshold are labeled accordingly. The 2-turn configuration violates CCM over most of the considered load-current range and is therefore excluded from the comparison.}
\vspace{-10pt}
\label{fig:power_loss}
\end{figure}

Over a 1~ms interval of an aggressive transient load profile, the adaptive converter exhibits 98 CCM violations without hysteresis and only 14 with hysteresis (Fig.~\ref{ccm_total_b}), substantially improving stability under worst-case workloads with only a small efficiency penalty (Fig.~\ref{fig:power_loss}). The remaining violations arise solely from a specific sequence of events: the low-inductance 2-turn state is latched (i.e., the sampled load current exceeds the entry threshold $I_{\mathrm{entry}}$), and within one ADC latency $T_{\mathrm{ADC}}=1$~\textmu s, the load current falls below the 2-turn CCM boundary $I_{\mathrm{CCM}}=\Delta I_L/2\approx1.2$~A. Only such rapid \emph{downward} load-current transitions can force DCM. Alternatively, slower decreases and all upward transitions are detected in time. Consequently, CCM is guaranteed provided
\begin{equation}
\left|\frac{dI}{dt}\right|_{\max} \;\leq\; \frac{I_{\mathrm{entry}}-I_{\mathrm{CCM}}}{T_{\mathrm{ADC}}}.
\label{eq:ccm_guarantee}
\end{equation}
For the considered stress profile (worst-case down-slew of $\sim$4.3~A/\textmu s), the nominal threshold ($I_{\mathrm{entry}}=2.41$~A) tolerates only 1.21~A/\textmu s and results in 98 CCM violations, whereas the 5~A hysteresis threshold tolerates 3.8~A/\textmu s and reduces the count to 14. Increasing $I_{\mathrm{entry}}$ to at least 5.5~A eliminates all violations, thereby deterministically guaranteeing CCM at a bounded efficiency penalty (Fig.~\ref{fig:power_loss}). While such deep sub-\textmu s downward load-current steps occur only in this deliberately aggressive stress profile, realistic HPC/AI workloads are unlikely to exhibit them.

\begin{figure}[tb]
\centering
\includegraphics[width=\columnwidth]{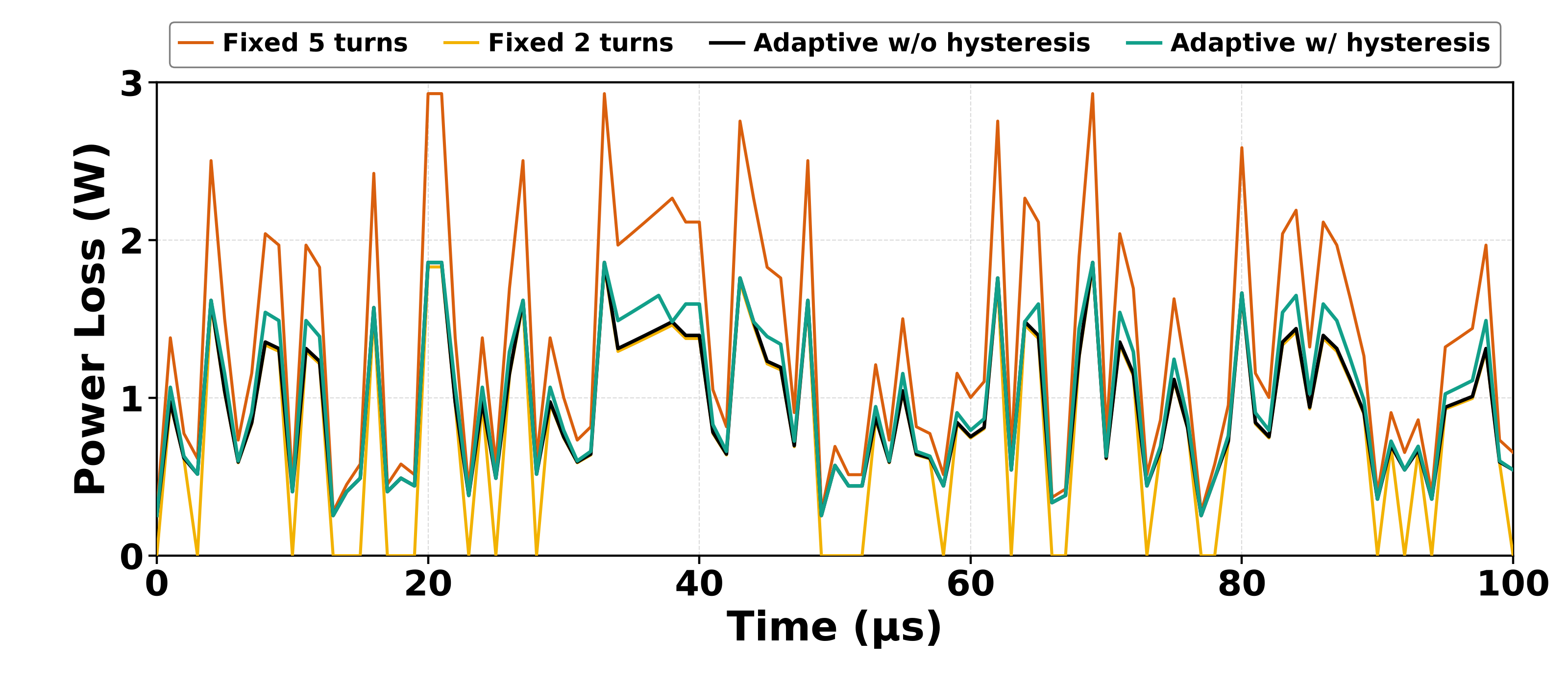}
\caption{Instantaneous power loss over a 100~\textmu s interval for the adaptive and fixed configurations, illustrating their dynamic behavior under varying load conditions.}
\vspace{-15pt}
\label{loss_and_averge_loss}
\end{figure}

Average power loss over the full 1~ms period for all considered configurations is summarized in Fig.~\ref{fig:power_loss}. The instantaneous power loss over a representative 100~\textmu s interval for some of the configurations is shown in Fig.~\ref{loss_and_averge_loss}. As expected, the hysteresis-based adaptive system exhibits higher loss than the nominal adaptive configuration, since it intentionally operates more often in suboptimal regimes to reduce repeated CCM violations; the hysteresis threshold should therefore be tuned to the application's burstiness and stability requirements.

Although the fixed 2T configuration achieves the lowest instantaneous power loss, it fails to maintain CCM when the load current falls below 1.2~A, thereby disqualifying it as a viable configuration. Similarly, the fixed 3T configuration achieves loss values comparable to those of the hysteresis-based adaptive system; however, it violates the ripple constraint even at the maximum switching frequency of 10~MHz, rendering it unsuitable for the specified design requirements.
In contrast, compared to the fixed 4T and 5T configurations, the proposed adaptive system provides significant loss reduction over the 1-ms bursty profile: (i) 18.8\% and 30.4\% for, respectively, hysteresis-based and nominal adaptive converters compared to the best valid fixed configuration (4 turns at 10~MHz), and (ii) 26.6\% and 32\% for, respectively, hysteresis-based and nominal adaptive converters compared to the worst valid configuration (5 turns at 10~MHz). These results highlight the effectiveness of the proposed adaptive approach in minimizing total power loss while satisfying design constraints across a wide range of operating conditions.


\vspace{-5pt}

\subsection{Circuit Validation}\label{sec:circuit_simulation}
\begin{figure}[tb]
\centering
\includegraphics[width=\columnwidth]{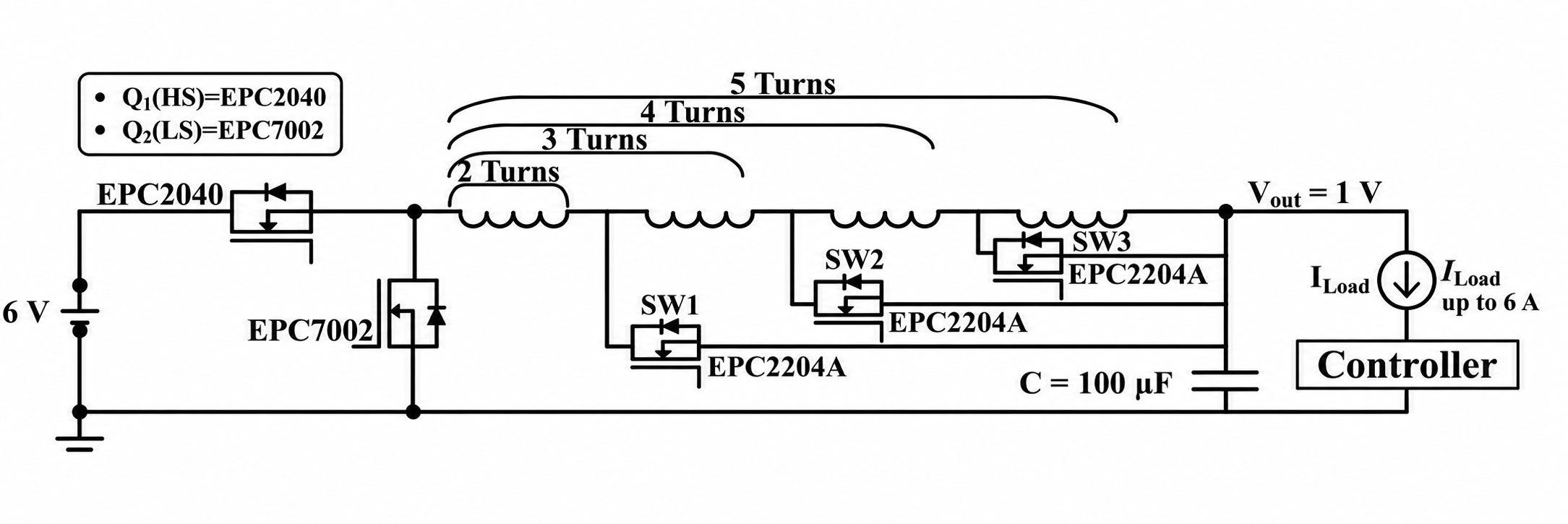}
\caption{Circuit-level simulation setup of the proposed adaptive converter, illustrating the reconfigurable inductor, power switches, and control circuitry.}
\vspace{-5pt}
\label{simulated_circuit}
\end{figure}

The proposed adaptive system, including the buck converter with the adaptive inductor and control circuitry is designed in Cadence Virtuoso and simulated using Spectre to validate its performance against the analytical models. 
The schematic of the system is shown in Fig.~\ref{simulated_circuit}. A 100~\textmu F capacitor is used to minimize the voltage ripple.

The adaptive inductor is implemented as a single tapped magnetic structure, modeled in Cadence as four series-connected inductive sections with intermediate nodes serving as taps. Compared to multiple discrete series-connected inductors, this implementation realizes all inductance states within a single magnetic component, sharing a common magnetic core and winding while avoiding duplicated magnetic components and their associated interconnects. This architecture is particularly well suited for integrated and package-embedded magnetics, where multiple discrete inductors would increase area, routing complexity, and fabrication effort. A network of three configuration switches selectively connects the taps to realize the 2-, 3-, 4-, and 5-turn operating modes. The electrical characteristics of each inductive section are derived from the preliminary FEM simulations and analytical models described in Section~\ref{Configurable-Converter}, with the resulting inductance and resistance values summarized in Table~\ref{tab:inductor_params}.

\begin{table}[tb]
\centering
\renewcommand{\arraystretch}{1.3}
\caption{Inductor Parameters for Different Turn Configurations From FEM Analysis.}
\begin{tabular}{|c|c|c|c|c|}
\hline
 & \textbf{2 turns} & \textbf{3 turns} & \textbf{4 turns} & \textbf{5 turns} \\
\hline
\textbf{Inductance} 
& 40.5 nH 
& 71.4 nH 
& 103.6 nH 
& 136.8 nH \\
\hline
\textbf{DC Resistance} 
& 27.28 m$\Omega$ 
& 39.6 m$\Omega$ 
& 51.1 m$\Omega$ 
& 61.9 m$\Omega$ \\
\hline
\end{tabular}
\vspace{-15pt}
\label{tab:inductor_params}
\end{table}

The controller is implemented in Verilog-A. The power overhead of the controller is expected to be negligible relative to the converter power stage due to its small footprint (40 logic gates and 24 \textmu W power consumption in TSMC 28 nm technology node). 
The power switching devices are modeled using Spectre device models, consistent with the assumptions adopted in the analytical framework. The control model employs two complementary strategies to regulate the output voltage under varying load conditions.

First, the converter duty cycle $D$ is dynamically adjusted as a function of the load current according to:

\begin{equation}
D = \frac{V_{out} + I_{Load}(R_L + R_{LS,on})}{V_{in} - I_{Load}(R_{HS,on} - R_{LS,on})},
\label{eq:duty_cycle}
\end{equation}
where $R_{HS,on}$, $R_{LS,on}$ are the on resistance of the respectively, high and low side switches, both extracted from the switch datasheets. 
Second, the switching frequency of the power devices is adaptively tuned based on results obtained from prior analytical optimization (see Table~\ref{tab:system_params}). 


\begin{figure}[tb]
\centering
\includegraphics[width=\columnwidth]{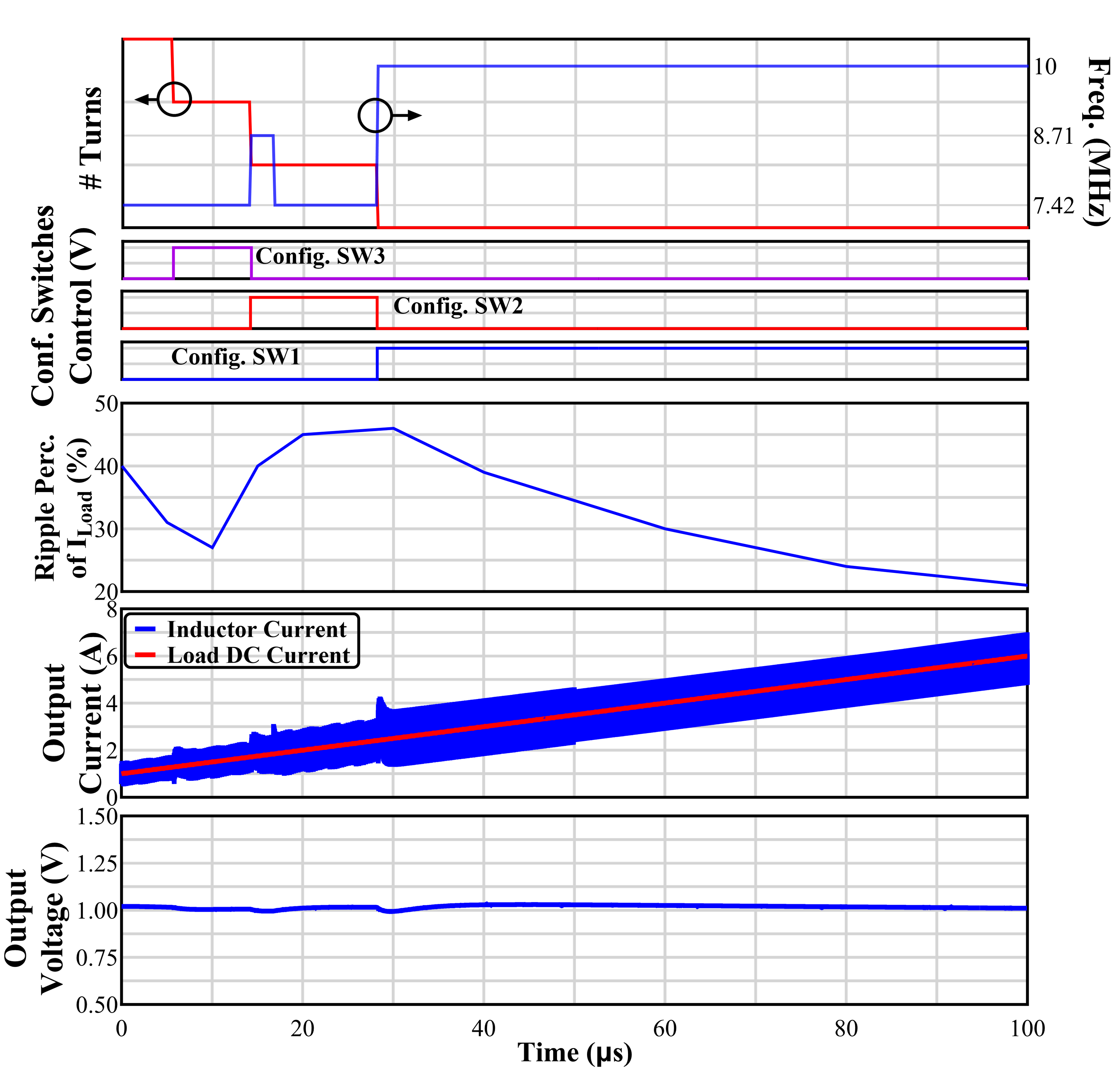}
\caption{Simulation results of the proposed adaptive converter, illustrating inductor and load currents, regulated output voltage, ripple percentage of load current, and the dynamic variation of inductor turns and switching frequency over a 100~\textmu s interval.}
\vspace{-10pt}
\label{sim_results}
\end{figure}

Transient simulation results over a 100~\textmu s interval for the overall adaptive system are presented in Fig.~\ref{sim_results}.
A ramp current profile, increasing from 1~A to 6~A over a 100~\textmu s interval, is applied as the DC load for the proposed converter. This setup enables a qualitative comparison with a conventional buck converter employing a fixed inductor and constant switching frequency, while utilizing identical switching devices and load capacitance. The converters are evaluated under a 6~V-to-1~V conversion. 
The output voltage is regulated within a $\pm3\%$ variation around 1~V across a load range of 1~A to 6~A. 

The inductor ripple current is also shown as a function of load, exhibiting CCM operation with an amplitude value of 0.4~A (40\%) at light load and of 1.26~A (21\%) at heavy loads. 

The performance of the non-hysteresis adaptive converter is compared with four conventional buck converters operating with fixed-turn inductors at a 10~MHz switching frequency under identical load conditions. The results are summarized in Table~\ref{tab:efficiency}. The proposed adaptive system demonstrates improved efficiency compared to the 3-, 4-, and 5-turn fixed-inductor buck converters, while the 3-turn configuration violates the ripple constraint (50\% of the load current) below 1.21 A load. Although the 2-turn fixed-inductor converter achieves higher efficiency than the adaptive system, it fails to maintain continuous conduction mode (CCM) and transitions to discontinuous conduction mode (DCM) at load currents below 1.05~A, and also violates the ripple constraint at loads below 2.2 A.

\begin{table}[tb]
\centering
\renewcommand{\arraystretch}{1.5}
\setlength{\tabcolsep}{5pt}
\caption{Simulated Efficiency Comparison of Adaptive and Fixed Converter Configurations.}

\label{tab:efficiency}
\begin{tabular}{|c|c|c|c|c|}
\hline
\textbf{Config.} & \textbf{Avg. Eff.} & \textbf{Max. Eff.} & \textbf{Min. Eff.} & \textbf{Max. Ripple} \\
\hline
Adaptive 
& 78.65 
& 85.88 (1.75 A) 
& 73.96 (6 A) 
& 46 (\%) \\
\hline
2T, 10 MHz 
& 80.17 
& 90.56 (1 A) 
& 75.22 (6 A) 
& \textcolor{red}{106 (\%)} \\
\hline
3T, 10 MHz 
& 77.11 
& 89.46 (1 A) 
& 71.41 (6 A) 
& \textcolor{orange}{60 (\%)} \\
\hline
4T, 10 MHz 
& 74.16 
& 88.24 (1 A) 
& 66.84 (6 A) 
& 42 (\%) \\
\hline
5T, 10 MHz 
& 72.03 
& 87.27 (1 A) 
& 65.43 (6 A) 
& 32 (\%) \\
\hline
\end{tabular}
\vspace{-10pt}
\end{table}

The circuit results agree with the analytical predictions (Fig.~\ref{generic_results_total_a}), reproducing the same trends and the same ranking of configurations; the differences are confined to the exact threshold currents---the 2-turn CCM exit shifts from 1.2 to 1.05~A, the ripple-violation points from 2.4~A to 2.2~A (2T) and 1.37 to 1.21~A (3T), and the adaptive peak-efficiency load from 1.9 to 1.75~A. These shifts stem from second-order effects captured only in simulation (device output capacitance and gate-charge dynamics, dead-time and body-diode conduction, finite output capacitance, and interconnect parasitics) and do not alter the overall trends or conclusions. The analytical and circuit evaluations also use different load excitations (a synthetic bursty profile vs.\ a deterministic ramp) and are thus not intended for one-to-one numerical comparison, yet both consistently demonstrate the superiority of the adaptive approach over fixed configurations.

\vspace{-5pt}

\subsection{Comparison with State-of-the-Art Converters}
\label{sec:sota}
Performance of the proposed and state-of-the-art converters is summarized in Table~\ref{tab:sota}. The converters in the table span various integration levels, converter architectures (including multiphase
buck~\cite{burton2014fivr, 10509290}, series-capacitor buck~\cite{shenoy2016scbc},
and recent hybrid/resonant switched-capacitor converters~\cite{ji2024ccp,
ren2025resc}), conversion ratios and load currents. Thus, a per-phase efficiency droop metric is used as a primary figure-of-merit: 
\begin{equation}
    \mathrm{FoM}_\mathrm{droop} = \frac{(\eta_{\mathrm{peak}}-\eta_{\mathrm{max-load}})N_\mathrm{phases}}{(I_{\mathrm{max-load}}-I_{\mathrm{peak-eff}})},
\end{equation}
\rr{where $\eta_{\mathrm{peak}}$ and $\eta_{\mathrm{max-load}}$ are the peak and
maximum-load efficiencies, $I_{\mathrm{peak-eff}}$ and $I_{\mathrm{max-load}}$ the
corresponding load currents, and $N_\mathrm{phases}$ normalizes the droop to a
per-phase basis so that designs with different phase counts and load spans compare
on equal terms.}

\rr{Two trends in Table~\ref{tab:sota} frame the result. First, the FoM correlates
strongly with form factor: roomy PCB designs achieve low droop
(0.35--2.44~pp/A~\cite{10509290,ren2025resc,ji2024ccp,shenoy2016scbc}) because
board area permits physically large, low-DCR, high-inductance windings whose
$I^2R$ and ripple remain small across the full load range---the lone PCB outlier
(10.47~pp/A~\cite{kandeel2023phasecount}) being a narrow-span design.
Such board-mounted discrete magnetics, however, cannot be readily integrated into a package
without increasing DCR and reducing inductance. Consequently, package- and interposer-integrated
regulators exhibit a substantially higher FoM (4.96~pp/A on
package~\cite{bellaredj2021solenoid}; 15.5--137~pp/A on
interposer~\cite{sturcken2013coupled,tien2015interposer}, the upper end inflated by
a small load span). 
The proposed planar air-core spiral contains no magnetic core and its footprint determined solely by its 200~\textmu m conductor, enabling direct integration into the package substrate alongside the switches.
Despite this compact implementation it achieves as low as 2.8~pp/A per phase (single-phase) and 2.72~pp/A per phase (four-phase)---PCB-comparable droop at package density, a distinction of integration level that a separate area column in the table would not capture. 
The lowest FoM in Table~\ref{tab:sota} ($\sim$1.6~(pp/A)$\cdot$phase) is achieved by Intel's FIVR through 1--16 phase shedding~\cite{burton2014fivr}, but for a much smaller conversion step (1.8-to-1.05~V versus the 6:1 ratio targeted here). This performance is inherently tied to that operating point. The 140~MHz switching frequency used to shrink the air-core inductors multiplies switching losses that already scale with the voltage step (overlap loss $\propto f(V_{in}-V_{out})$ and output-capacitance loss $\propto f(V_{in}-V_{out})^2$), making the same approach prohibitive for efficient 6:1 conversion. Moreover, phase shedding maintains each active phase near its peak-efficiency current by increasing the number of phases as the supported load range expands. In addition, because phase shedding activates the fewest phases at the lightest loads, it forfeits interleaved ripple cancellation exactly where the low current makes the relative ripple, and hence the CCM margin, most critical.
Under a fixed relative-ripple constraint, reducing the per-phase current requires a proportional increase in per-phase inductance at a fixed switching frequency and conversion ratio. A package-level implementation over the same load range would therefore require multiple high-inductance magnetic elements, whose realization entails substantially greater footprint and/or vertical volume, or increased DCR and AC loss. In~\cite{burton2014fivr}, this penalty is suppressed through the very high switching frequency and small conversion step; such frequencies are generally impractical for package-integrated regulators, whereas implementing 6:1 conversion directly on-chip is likewise impractical. The proposed converter instead reuses the same magnetic structure, increasing its effective inductance at light load and reconfiguring it for lower inductance and DCR at heavy load. It thereby maintains high efficiency across the load range at an order-of-magnitude lower switching frequency without the magnetic replication required by fine-grained phase shedding. Conventional phase shedding remains complementary.
}

\begin{table*}[t]
\centering
\renewcommand{\arraystretch}{1.3}
\setlength{\tabcolsep}{5pt}
\caption{Comparison with State-of-the-Art Integrated Voltage Regulators.}
\label{tab:sota}
\resizebox{\textwidth}{!}{%
\begin{tabular}{l|l|c|c|c|c|c|c|c|c}
\hline
\textbf{Work} & \textbf{VR Topology / Magnetics} & \textbf{\# Phases} & \textbf{$V_{in}\!\to\!V_{out}$ (V)} & \textbf{$f_s$ (MHz)} & \textbf{Peak $\eta$} & \textbf{High-load $\eta$} & \textbf{FoM} & \textbf{Results} & \textbf{Integration} \\
\hline
Shenoy, TPEL'16~\cite{shenoy2016scbc} & Buck / series-cap buck & 2 & $12\!\to\!1.2$ & 1--3 & 90\% @1 A& 79\% @10\,A & 2.44 pp/A & Measured & PCB \\
Gan, APEC'24~\cite{10509290} & IBC + buck (VPD) & 24 & $48\!\to\!1$ & 1.2 & 87\% @300 A & 79\% @850\,A & 0.35 pp/A & Measured & PCB \\
Ji, JSSC'24~\cite{ji2024ccp} & Hybrid flying-cap buck & 1 & $12\!\to\!1.2$ & 1 & 93.7\% @1 A & 84.8\% @5\,A & 2.23 pp/A & Measured & PCB \\
Ren, JSSC'25~\cite{ren2025resc} & IBC + resonant-SC & 3 & $12\!\to\!1$ & 9 & 89.3\% @3.9 A& $\sim$86\% @12\,A & 1.22 pp/A & Measured & PCB \\
Kandeel, TPEL'23~\cite{kandeel2023phasecount} & Buck & 2 & $4.5\!\to\!1.8$ & 20 & $\sim$90\% @1.09 A& $\sim$80\% @3\,A & 10.47 pp/A & Measured & PCB \\
Sturcken, JSSC'13~\cite{sturcken2013coupled} & Multiphase coupled buck & 8 & $1.8\!\to\!1.2$ & 100 & 75\% @3.2\,A & 69\% @6.3\,A & 15.48 pp/A & Measured & Interposer \\
Tien, VLSI'15~\cite{tien2015interposer} & Multiphase buck & 8 & $1.66\!\to\!0.83$ & 150 & 82\% @0.39\,A & $\sim$75\% @0.8\,A & 136.6 pp/A & Measured & Interposer \\
Burton, APEC'14~\cite{burton2014fivr} & Buck with phase shedding & 2-16 & $1.8\!\to\!1.05$ & 140 & 90.5\% @6 A & $\sim$89.5\%$^a$ @16\,A & 1.6 pp/A & Measured & On-die \\
Burton, APEC'14~\cite{burton2014fivr} & Buck with phase shedding & 2-4 & $1.8\!\to\!1.05$ & 140 & 89.5\% @3 A & 80\%$^a$ @16\,A & 2.92 pp/A & Measured & On-die \\
Bellaredj, TCPMT'21~\cite{bellaredj2021solenoid} & Buck & 4 & $5\!\to\!1$ & 100 & $\sim$74\% @5 A& 67.8\% @10\,A & 4.96 pp/A & Simulated & Package \\
\textbf{This work} (1-phase) & Reconfigurable buck & 1 & $6\!\to\!1$ & 7.4--10 & 85.9\% @1.75\,A & 74.0\% @6\,A & 2.8 pp/A & Simulated & Package \\
\textbf{This work} (4-phase) & Reconfigurable buck & 4 & $6\!\to\!1$ & 7.4--10 & 86.84\% @7A & 83.44\% @12\,A & 2.72 pp/A & Simulated & Package \\
\hline
\multicolumn{10}{l}{\ssmall $^a$Derived based on Figure 4(b) in ~\cite{burton2014fivr}.}
\end{tabular}}
\vspace{-15pt}
\end{table*}

\begin{figure}[tb]
\centerline{\includegraphics[width=\columnwidth]{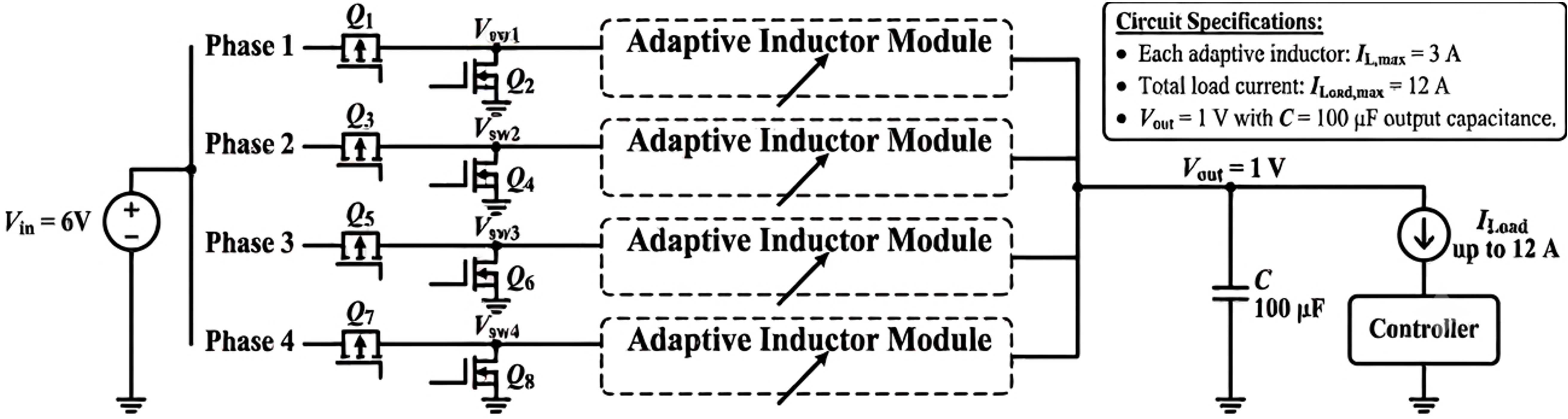}}
\caption{Four-phase adaptive-inductor buck converter designed for a controlled
comparison with~\cite{bellaredj2021solenoid}: each phase employs a reconfigurable
(adaptive) inductor rated to $I_{L,\max}=3$~A, delivering a total load current of up
to 12~A at $V_{out}=1$~V from a 6~V input.}
\vspace{-15pt}
\label{fig:two_phase}
\end{figure}

The comparison with SOTA in Table~\ref{tab:sota} provides quantitative context rather than a definitive ranking. Although the FoM normalizes the efficiency variation by phase count and load span, conversion ratio, integration level, switching frequency, and magnetic implementation also influence both efficiency and the FoM.
The package-embedded four-phase solenoid IVR of Bellaredj~\emph{et al.}~\cite{bellaredj2021solenoid} is the most directly comparable reference because it targets package integration, uses the same analytical efficiency modeling framework, and is evaluated over a comparable load range. This provides the fairest apples-to-apples comparison of the proposed adaptive architecture.

To address the phase difference between the two converters, a four-phase variant of the proposed converter (Fig.~\ref{fig:two_phase}) is designed at the same 6:1 conversion ratio and 7.4--10~MHz switching frequency, with each phase delivering up to 3~A (12~A total). A peak efficiency of 86.84\% is achieved at 7~A, decreasing to 83.44\% at 12~A. By comparison, the closest operating point reported in~\cite{bellaredj2021solenoid} (5:1 conversion ratio) peaks at $\sim$74\% (5~A) and decreases to $\sim$67.8\% (10~A), corresponding to a FoM of 4.96~(pp/A) per phase versus 2.72~(pp/A) per phase for the proposed converter. This nearly twofold reduction in the normalized efficiency droop is achieved despite the larger conversion ratio, demonstrating the benefit of coordinated inductance and switching-frequency adaptation under identical modeling assumptions. Because both converters are evaluated using the same analytical framework, systematic errors in the absolute efficiency largely cancel, making the relative FoM comparison substantially more robust than cross-paper efficiency comparisons. Furthermore, because the proposed air-core spiral (200~\textmu m conductor, $\leq$50\% ripple, 7.4--10~MHz) intentionally retains a significant winding $I^2R$ contribution (up to 30\% of the total loss at heavy load), the improvement directly reflects the effectiveness of the adaptive technique. The same adaptation strategy is expected to remain applicable to lower-DCR magnetic structures and higher-phase-count converters.
\vspace{-5pt}

\section{Conclusion}
\label{Conclusion}
\rr{A digitally controlled adaptive power conversion framework is presented, in which both the inductor turns and the switching frequency are dynamically adjusted to maintain high efficiency across varying load conditions. Through joint optimization of inductance and switching frequency, conduction and switching losses are balanced while the ripple and CCM constraints are satisfied.}
A complete design methodology combining analytical modeling, FEM-based characterization, and optimization is developed, with real-time implementation achieved through a lookup-table-based controller. A hysteresis mechanism is introduced to improve robustness against transient CCM violations; the associated entry-margin condition~\eqref{eq:ccm_guarantee} further shows that continuous conduction can be
guaranteed deterministically for a bounded load slew rate, at a small efficiency cost.
Over the 1--6~A load range, an average efficiency improvement of 4.5 percentage points is demonstrated relative to the best valid fixed configuration (and 6.1 pp relative to the least-efficient valid configuration). Over a realistic bursty load profile, the nominal adaptive system is shown to reduce the average power loss by 30.4\% relative to the best valid fixed configuration (4 turns at 10~MHz) and by 32.0\% relative to the worst valid configuration (5 turns at 10~MHz), while the hysteresis-based variant achieves 18.8\% and 26.6\% reductions, respectively, trading a modest increase in loss for substantially fewer transient CCM violations. These results highlight the effectiveness of coordinated inductance and switching frequency adaptation for next-generation power delivery systems operating under dynamic load profiles.

\vspace{-5pt}

\section*{Acknowledgment}
This work was supported by the Center for Heterogeneous Integration of Micro Electronic Systems (CHIMES), one of seven centers in Joint University Microelectronics Program (JUMP) 2.0, a Semiconductor Research Corporation (SRC) program sponsored by the Defense Advance Research Project Agency (DARPA).

\vspace{-5pt}

\bibliographystyle{IEEEtran}
\bibliography{Refs}
\end{document}